\documentclass[preprint,amsmath,amssymb,prb,floatfix,superscriptaddress]{revtex4}

\usepackage[dvips]{graphicx}
\usepackage{dcolumn}
\usepackage{bm}
\usepackage{mathrsfs}

\begin{document}
\title{Long wavelength magnetic and magnetoelectric excitations in the ferroelectric antiferromagnet BiFeO$_3$}
\author{D. Talbayev}
 \email{diyar.talbayev@yale.edu}
 \affiliation{Department of Chemistry, Yale University, PO Box 208107, New Haven, CT 06520-8107, USA}
\author{S.A. Trugman}
 \affiliation{Center for Integrated Nanotechnologies, MS K771, Los Alamos National Laboratory, Los Alamos, NM 87545, USA}
\author{Seongsu Lee}
 \affiliation{Rutgers Center for Emergent Materials and Department of Physics and Astronomy, Rutgers University, 136 Frelinghuysen Rd., Piscataway, NJ 08854, USA}
\author{Hee Taek Yi}
 \affiliation{Rutgers Center for Emergent Materials and Department of Physics and Astronomy, Rutgers University, 136 Frelinghuysen Rd., Piscataway, NJ 08854, USA}
\author{S.-W. Cheong} 
 \affiliation{Rutgers Center for Emergent Materials and Department of Physics and Astronomy, Rutgers University, 136 Frelinghuysen Rd., Piscataway, NJ 08854, USA}
\author{A. J. Taylor} 
 \affiliation{Center for Integrated Nanotechnologies, MS K771, Los Alamos National Laboratory, Los Alamos, NM 87545, USA}
\date{\today}

\newcommand{\cm}{\:\mathrm{cm}^{-1}}
\newcommand{\T}{\:\mathrm{T}}
\newcommand{\mc}{\:\mu\mathrm{m}}
\newcommand{\ve}{\varepsilon}
\newcommand{\dg}{^\mathtt{o}}

\begin{abstract}
We present a terahertz spectroscopic study of magnetic excitations in ferroelectric antiferromagnet BiFeO$_3$. We interpret the observed spectrum of long-wavelength magnetic resonance modes in terms of the normal modes of the material's cycloidal antiferromagnetic structure. We find that the modulated Dzyaloshinski-Moriya interaction leads to a splitting of the out-of-plane resonance modes. We also assign one of the observed absorption lines to an electromagnon excitation that results from the magnetoelectric coupling between the ferroelectric polarization and the cycloidal magnetic structure of BiFeO$_3$.
\end{abstract}

\maketitle

Ferroelectric antiferromagnet BiFeO$_3$ (BFO) combines ferroelectricity\cite{smith:70} with an antiferromagnetic order at room temperature\cite{sosnowska:4835}. Materials that display both magnetic and ferroelectric orders, multiferroics, hold a promise of advanced devices that exploit the magnetoelectric (ME) effect, e.g., the manipulation of magnetic state by electric field\cite{wood:303, fiebig:123, cheong:13, ramesh:21}. BFO is of considerable interest in this respect, as it is a high-temperature ferroelectric\cite{wang:1719, shvartsman:172115, lebeugle:22907} ($T_c\approx$1100 K) with a large ferroelectric dipole moment $\sim$100 $\mu$C/cm$^2$. A control of its magnetic state by voltage has been demostrated both in bulk and in thin film BFO\cite{lebeugle:227602, lee:192906, zhao:823, chu:478}. At room temperature, bulk crystalline BFO adopts a G-type antiferromagnetic structure ($T_N$=640 K) with a cycloidal spin arrangement with a long modulation period of $\lambda\approx620$ \AA\cite{sosnowska:4835, lee:192906, lebeugle:227602}. The cycloidal modulation of the antiferromagnetic vector in BFO results from ME interaction between the ferroelectric polarization and magnetic moments in the form of the Lifshitz invariant\cite{kadomtseva:571}. 

Spectroscopic studies of magnetic and lattice excitations have long been used to gain insight in the nature of both magnetism and ferroelectricity\cite{kamba:024403, lobo:172105, scott:83}. Magnetic excitations (magnons), in particular, are extremely sensitive to the microscopic magnetic interactions in a material. In antiferromagnets, magnon resonance frequencies and dispersion are determined by the strength of the antiferromagnetic exchange and by magnetic anisotropy\cite{foner}. In multiferroic manganites, far-infrared spectroscopic studies of magnetic excitations helped unravel the underlying microscopic ME interactions that were found to mix magnons with phonons\cite{valdez:047203, pimenov:014438}. In this article, we present a terahertz (THz) spectroscopic study of long-wavelength magnetic excitations in BFO. We find several distinct magnetic-dipole absorption resonances that are interpreted in terms of the normal modes of the antiferromagnetic cycloid. The selection rules for magnetic-dipole transitions show that the main observed magnetic resonances are not uniform modes, but those with the wavevector of the cycloid. A comparison with a calculation of magnon frequencies in BFO\cite{desousa:012406} shows that the antiferromagnetic exchange and the Lifshitz invariant terms in the free energy are insufficient to describe the magnetic excitation spectrum. One important additional term in the free energy is the modulated Dzyaloshinksi-Moriya (DM) exchange interaction that lends a weak local ferromagnetic moment to the cycloid, although the macroscopic magnetization remains zero\cite{kadomtseva:571,lee:100101}. We conjecture that the modulated DM exchange also leads to a characteristic splitting of non-uniform normal modes that is prominent in our measured spectra. We also find a resonant absorption mode at $18.4\cm$ that cannot be accounted for by magnetic-dipole-active excitations of the cycloid but obeys the selection rule for an electromagnon\cite{desousa:012406}. This electromagnon excitation couples to the $ac$ electric field of the THz wave and results from magnetoelectric coupling between the cycloid and the $E_g$ optical phonon in BFO.

Earlier Raman investigations of magnetic modes in BFO reported a series of magnetic resonance lines that was described as uniform and higher-order non-uniform magnetic modes of the cycloid\cite{singh:252203, cazayous:037601,rovillain:180411}. In some cases (Cazayous et al.\cite{cazayous:037601}), the relative strength of the observed higher-order magnetic Raman modes compared to the uniform modes was attributed to their electromagnon character. We observe a lower number of magnetic resonance lines than in the Raman investigations. Different selection rules between Raman and THz experiments may explain the disparity. As we show in this article, either uniform or higher-order magnetic modes of the cycloid are silent for certain polarizations of THz light.

In bulk crystals, BFO adopts a rhombohedral $R3c$ structure, which can be obtained from the cubic perosvkite structure by elongating the cubic unit cell ($a\approx 3.96$ \AA, $\alpha\approx89.4\dg$) along the [111] diagonal\cite{sosnowska:4835}, which also corresponds to the direction of the ferroelectric polarization $\bm{P}$. The antiferromagnetic cycloid wavevector $\bm{q}$ belongs to the plane perpendicular to the ferroelectric polarization and possesses 3 equivalent directions in that plane: [10-1], [1-10], and [01-1] (Fig.~\ref{fig:tempfreq}). The plane of the cycloid is defined\cite{sosnowska:4835} by vectors $\bm{P}$ and $\bm{q}$, within which the antiferromagnetic vector $\bm{L}=\bm{M_1}-\bm{M_2}$ slowly rotates with the period $\lambda$. $\bm{M_1}$ and $\bm{M_2}$ are magnetic moments on nearest neighbor Fe ions. The magnitude of $\bm{q}=2\pi/\lambda$ is determined by the competition between the antiferromagnetic exchange and the ME interaction in the form of the Lifshitz invariant in the magnetic free energy
\begin{equation}
F=A\sum_i{(\nabla L_i)^2}-\gamma\bm{P}\cdot\left[\bm{L}(\nabla\cdot\bm{L})+\bm{L}\times(\nabla\times\bm{L})\right],
\label{eq:free1}
\end{equation}
where $A$ is the exchange stiffness constant and $\gamma$ is the ME coefficient. The minimum of the free energy in Eq. (\ref{eq:free1}) occurs at the wave vector\cite{kadomtseva:571, desousa:012406} $q=\gamma P/2A$, which allows the estimate of ME coefficient $\gamma=10^5$ erg/C using the known values\cite{kadomtseva:571} of $\lambda$ and $A=3\times10^{-7}$ erg/cm. The corresponding cycloidal magnetic structure is described by a harmonic spatial dependence of the antiferromagnetic vector $\bm{L}=L_0 \left[ \cos(qx)\bm{\hat{z}} + \sin(qx)\bm{\hat{x}}\right]$ along the direction of vector $\bm{q}$ (Fig.~\ref{fig:tempfreq}). At low temperatures ($T\leq77$ K), the cycloid was found to develop strong anharmonicity due to uniaxial magnetic anistropy with an easy $c$ axis\cite{zalessky:547, kadomtseva:571}. The anisotropy causes the spins to orient preferentially within a small angle with the $c$ axis for most of the cycloid period, as was deduced from asymmetric shapes of nuclear magnetic resonance lines\cite{zalessky:547}.

\begin{figure}[ht]
\begin{center}
\includegraphics[width=4in]{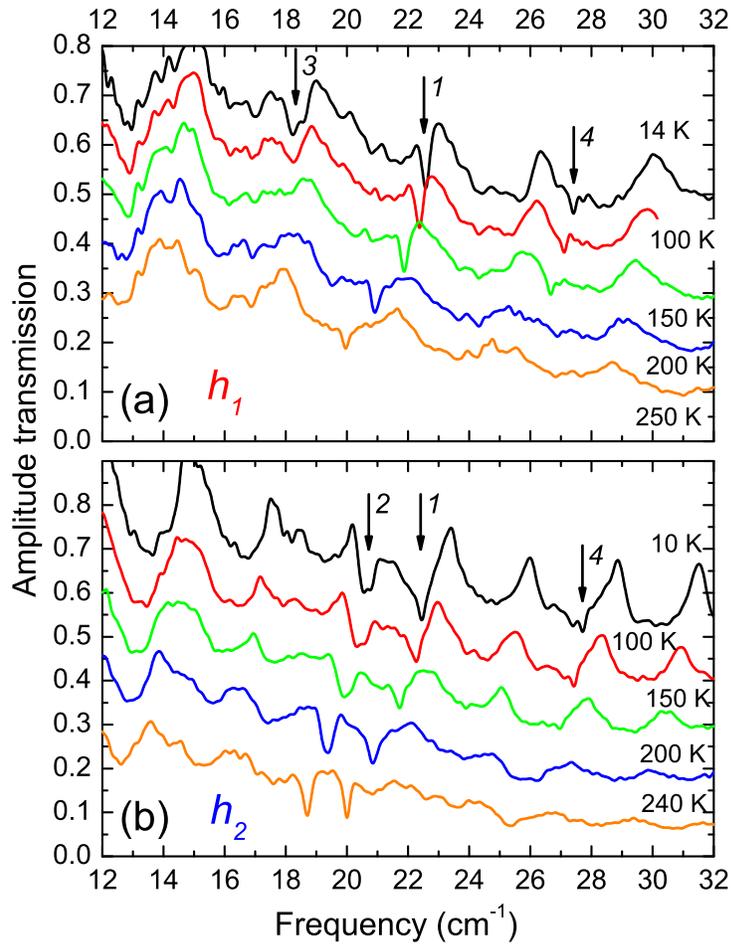}
\caption{\label{fig:espectra}(Color online) THz amplitude transmission spectra of BiFeO$_3$ at different temperatures. (a) Polarization $h_1$. (b) Polarization $h_2$. The spectra are offset vertically for clarity. Vertical arrows indicate the observed magnetic resonance lines.}
\end{center}
\end{figure}

\begin{figure}[ht]
\begin{center}
\includegraphics[width=4in]{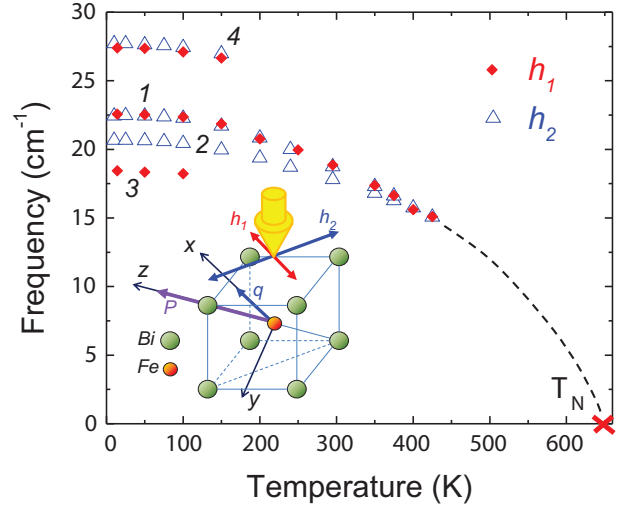}
\caption{\label{fig:tempfreq}(Color online) Temperature dependence of the frequencies of all observed magnetic resonance modes. The resonance labels are the same as in Fig.~\ref{fig:espectra}. Inset: geometry of the BFO crystal and of the THz transmission measurement.}
\end{center}
\end{figure}

The bulk BFO crystal used in this work was grown using Bi$_2$O$_3$ flux. Our measurements were carried out on a 220$\mu$m-thick single crystal with (001)$_{cubic}$ orientation mounted on a 3 mm aperture. The crystal was found to consist of a single ferroelectric domain in THz wave emission experiments\cite{talbayev:212906}, in which the orientation of the ferroelectric $c$ axis was also determined. THz transmission of the crystal was measured in two different polarizations ($h_1$ and $h_2$) using a home-built time-domain THz spectrometer based on photoconductive switches used as emitter and receiver of the THz wave. In the inset of Fig.~\ref{fig:tempfreq}, $h_1$ and $h_2$ designate the $ac$ magnetic field of the THz wave incident on the (001)$_{cubic}$ face of the BFO sample mounted on the cold finger of a He flow cryostat. Complex amplitude transmission of the sample was measured in time domain by using an empty aperture of the same size as a reference. Fig.~\ref{fig:espectra} shows the amplitude of the complex transmission in the two polarizations at various temperatures. Prominent absorption resonances indicated by vertical arrows (and labeled with numbers 1, 2, and 3) can be distiguished in the spectra. The oscillation of the transmission amplitude with a 1.5-2~$\cm$ period is due to a Fabri-Perot effect between the front and back surfaces of the sample. The $h_1$- and $h_2$-polarization spectra differ sharply, as the resonance 2 (at 20.5~$\cm$ at 10 K) is completely absent from the $h_1$ spectrum and the resonance 3 (at 18.4~$\cm$ at 14 K) is completely absent from the $h_2$ spectrum. The resonance 1 (at 22.5~$\cm$ at 10K) is observed and shows a similar strength in both polarizations.

We assign the resonances 1 and 2 to magnetic-dipole transitions corresponding to the normal modes of magnetic motion of the cycloid (antiferromagnetic resonance, AFMR), as their measured frequency is considerably lower than the frequency of the lowest optical phonon in BFO\cite{kamba:024403, lobo:172105} - the $E_g$ mode at $74\cm$ (at 5 K). Magnetic excitations at similar frequencies to our resonances have also been observed in single crystals using Raman scattering\cite{singh:252203, cazayous:037601,rovillain:180411} and in submillimeter wave spectroscopy on BFO ceramics\cite{komandin:734}. Fig.~\ref{fig:tempfreq} displays a softening of resonance frequencies with increasing temperature, which is typical of AFMR frequencies that are expected to reach zero a $T_N$ (dashed line in Fig.~\ref{fig:tempfreq}). The strong polarization dependence of the measured spectra suggests that the magnetic state of the sample consists of a single magnetic domain out of the 3 energetically equivalent magnetic domains corresponding to the 3 possible directions of the cycloid wavevector $\bm{q}$. Since the direction of $\bm{q}$ also determines the plane of the cycloid, we argue that the actual orientation of $\bm{q}$ in our sample is that shown in the inset of Fig.~\ref{fig:tempfreq} with $\bm{q}$ parallel to the magnetic field of the THz wave in $h_1$ polarization. As it is the only orientation of $\bm{q}$ that allows the magnetic field of the incident THz wave in the two polarizations to be either along or perpendicular to the $\bm{q}$ direction, the transmision spectra for this magnetic domain are expected to exhibit the sharpest difference between polarizations $h_1$ and $h_2$.

As evident from Fig.~\ref{fig:espectra}, resonance 3 disappears at 150 K, while resonances 1 and 2 persist until about 400 K, when they too become unobservable. We suggest that the disappearance of resonance 3 results from a phase transition occurring around 140 K \cite{singh:252203, scott:322203, cazayous:037601, xu:134425} and associated with a spin reorientation. The normal modes of the cycloid described by the free energy in Eq. (\ref{eq:free1}) where calculated by de Sousa and Moore\cite{desousa:012406}, who found that in addition to zero-wavevector magnons, the magnons at integer multiples of the cycloid wavevector $\bm{q}$ can couple to the THz wave due to the periodicity of the static magnetic structure and magnon zone folding. We will now use those normal modes together with magnetic dipole selection rules to try and assign the resonances 1 and 2 to specific magnons of the cycloid.

\begin{figure}[ht]
\begin{center}
\includegraphics[width=4in]{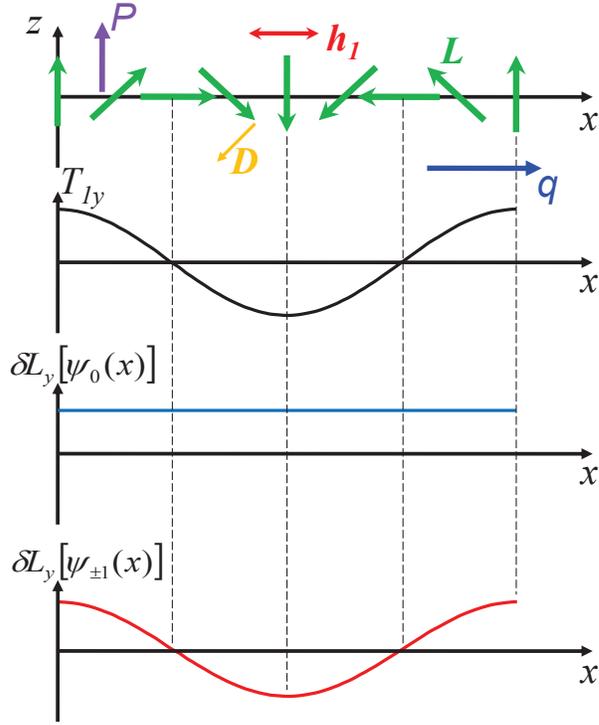}
\caption{\label{fig:cycloidtorque}(Color online) Antiferromagnetic cycloid (vector $\bm{L}$)  and coupling of torque $T$ to cycloid motion (vector $\bm{\delta L}$) in modes $\psi_{0}$, $\psi_{\pm 1}$ in $h_1$ polarization. Also shown are the ferroelectric polarization $\bm{P}$, the cycloid propagation vector $\bm{q}$, and vector $\bm{\hat{D}}$ of cycloid motion in mode $\phi_0$.}
\end{center}
\end{figure}

We start by briefly summarizing the findings of de Sousa and Moore, who parametrized the small motions of the antiferromagnetic vector $\bm{L}$ in the normal modes of the cycloid as
\begin{equation}
\delta \bm{L}=\left(\phi(r)\hat{\bm{D}}(x)+\psi(r)\hat{\bm{y}}\right)e^{-i\omega t},
\label{eq:lmotion}
\end{equation}
where vector $\hat{\bm{D}}(x)=\cos(qx)\hat{\bm{x}}-\sin(qx)\hat{\bm{z}}$ belongs to the cycloid plane and is transverse to vector $\bm{L}$ (Fig.~\ref{fig:cycloidtorque}). Equation (\ref{eq:lmotion}) shows that the normal modes separate into in-plane cyclon modes ($\phi$) and out-of-plane modes ($\psi$). The zero-wavevector (uniform) cyclon mode $\phi_0$ is gapless, i.e., has zero eigenfrequency, while the uniform out-of-plane mode $\psi_0$ has an eigenfrequency $\omega_{0}^{\psi}=\gamma P/\sqrt{2A}$. In addition to the uniform modes of motion, non-uniform $\phi_n, \psi_n\sim e^{inqx}$ modes appear in the THz spectrum of BFO due to the periodic nature of the cycloid (magnon zone folding). The frequencies of non-uniform modes form the series $\omega_{n}^{\psi}=\gamma (P/\sqrt{2A}) \sqrt{n^2+1}$ and $\omega_{n}^{\phi} = \gamma (P/\sqrt{2A}) \left|n\right|$ with integer $n\neq 0$.

To determine the magnetic-dipole selection rules for the $\phi_n$ and $\psi_n$ modes, we consider the torque of the $ac$ magnetic field of the THz wave on the antiferromagnetic vector $\bm{L}$: $\bm{T}=\bm{L}\times\bm{h}$. In polarization $h_1$, the $ac$ field $\bm{h_1}$ belongs to the $x-z$ plane and is parallel to direction of vector $\bm{q}||\bm{\hat{x}}$ (Fig.~\ref{fig:tempfreq} inset). The corresponding torque $T_1$ possesses non-zero components only along the $\bm{\hat{y}}$ direction, and $T_{1y}(x)$ varies sinusoidally with $x$ (Fig.~\ref{fig:cycloidtorque}) for the harmonic cycloid. The absence of $\bm{\hat{x}}$ and $\bm{\hat{z}}$ components in $T_1$ prohibits the coupling to the in-plane $\phi_n$ modes in $h_1$ polarization for any $n$. The sinusoidal variation of $T_{1y}(x)$ determines which of the $\psi_n$ modes couple to the $h_1$ polarization. For example, as the product $\delta L_y\left[\psi_0(x)\right] \cdot T_{1y}(x)$ averages to zero over a cycloid period, the coupling of $h_1$ polarization to the $\psi_0$ mode is prohibited. In fact, the quantity $\delta L_y\left[\psi_n (x)\right] \cdot T_{1y}(x)$ averages to a non-zero value only for $n=\pm 1$, as shown in Fig.~\ref{fig:cycloidtorque}, which means that only the $\psi_{\pm 1}$ modes can be excited in $h_1$ polarization in a harmonic cycloid. Our measured spectra contain only one absorption line in the broad temperature range in $h_1$ polarization (with the exception of resonance line 3, which is only observed below 150 K and will be discussed in more detail below). This agrees with the selection rule that we described, as we expect the frequencies of the modes with $n=\pm 1$ to be the same. Thus, we assign the resonance line 1 as the $\psi_{\pm 1}$ modes in $h_1$ polarization.

In $h_2$ polarization, the $ac$ magnetic field has components along the $\bm{\hat{y}}$ and $\bm{\hat{z}}$ directions: $h_2^y$ and $h_2^z$. Considering the components of the torque of field $h_2^y$, we find that $\bm{T}=\bm{L} \times \bm{h_2^y} \propto \bm{\hat{D}}$. Accordingly, the only mode that can be excited by the field $h_2^y$ is the uniform $\phi_0$ mode with zero frequency in the model described by Eq. (\ref{eq:free1}). Such low frequency makes the mode unobservable in our spectra.
The torque of the field $h_2^z$ has only components along the $\bm{\hat{y}}$ direction and only excites the $\psi_{\pm 1}$ modes. The measured spectra display two prominent absorption lines in the $h_2$ polarization. Since the $\psi_{\pm 1}$ modes have the same frequency in the de Sousa-Moore description\cite{desousa:012406}, the origin of the observed doublet needs to be explained. Possible candidates for additional absorption lines include the higher order $\psi_n$ modes with $n>1$ that become allowed for an anharmonic cycloid. When the anharmonicity is caused by easy-axis anisotropy, only odd integers $n=1,3, ...$ are allowed in the series $L_z=L_0 \sum_n b_n\cos(nqx)$ describing the anharmonic cycloid (the $L_x$ components are set by the condition $L_0^2=L_z^2+L_x^2$). The strength of such higher order absorption with $|n|\geq 3$ is expected to be considerably lower than that of the $\psi_{\pm 1}$ absorption, while our spectra show an almost identical absoprtion strength for the components of the doublet. In addition, frequency separation between $\psi_{\pm 1}$ and $\psi_{\pm 3}$ modes at the lowest temperatures is expected to be\cite{singh:252203,cazayous:037601} $\sim7.5-8\cm$, which is much higher than the observed separation of $1.7\cm$. On these grounds, we exclude the higher order modes from being a part of the doublet.

Since the two lines of the doublet have almost identical oscillator strength, we propose that they represent a splitting between $\psi_{+1}$ and $\psi_{-1}$ modes. To achieve such splitting, additional terms are needed in Eq. (\ref{eq:free1}). One of the terms that have been omitted from Eq. (\ref{eq:free1}) is a modulated DM term of the form\cite{kadomtseva:571, ruette:064114}
\begin{equation}
F_{DM}=DP_z(M_yL_x-M_xL_y),
\label{eq:dmmod}
\end{equation}
where $D$ is the DM constant. The DM term leads to a local canting of antiferromagnetic sublattices that results in a local magnetization along the $\bm{\hat{y}}$ direction $M_y=M_{1y}+M_{2y}$ that is modulated with the same wavevector $\bm{q}$ as the cycloid and does not give rise to a macroscopic magnetic moment\cite{kadomtseva:571}. Since the weak magnetization $M_y$ is modulated with the same wavevector as the antiferromagnetic cycloid, we conjecture that it is the DM term of Eq. (\ref{eq:dmmod}) that causes the $\psi_{+1}$/$\psi_{-1}$ splitting.

The $\psi_{+1}$/$\psi_{-1}$ splitting is not observed in $h_1$ polarization because the absolute value of the torque $T_{1y}(x)$ reaches its maxima at points along the cycloid direction $x$ where the vector $\bm{L}$ aligns along the $z$ direction and where $L_x=L_y=0$ (Fig.~\ref{fig:cycloidtorque}). These are the points of the strongest coupling of the $h_1$-polarized wave to the $\psi_{\pm 1}$ modes of the cycloid. At the same time, these are the points where the canting of antiferromagnetic sublattices is zero, according to Eq. (\ref{eq:dmmod}). The strongest sublattice canting happens at the points where the torque $T_{1y}(x)$ vanishes (Eq. (\ref{eq:dmmod}) and Fig.~\ref{fig:cycloidtorque}). This makes the wave of $h_1$ polarization insensitive to the $\psi_{+1}$/$\psi_{-1}$ splitting. By contrast, the absolute value of the torque of the field $h_2^z$ is maximum at the same points where the sublattice canting is the highest, which renders the $\psi_{+1}$/$\psi_{-1}$ splitting observable. In the description of de Sousa and Moore, all $\psi_{\pm 1}$ modes are degenerate and possess the same resonance frequency. Our measurements and analysis show that the introduction of the DM term of Eq. (\ref{eq:dmmod}) lifts the degeneracy and splits the $\psi_{\pm 1}$ modes into at least three groups - the doublet observed in $h_2$ polarization and the single mode observed in $h_1$ polarization. Perhaps by accident, one line of the doublet and the singlet display very similar frequencies (Figs.~\ref{fig:espectra} and~\ref{fig:tempfreq}). A calculation of magnetic resonance modes taking account of the DM term is needed to confirm and completely describe the observed splitting.

A closer look at the spectra in Fig.~\ref{fig:espectra} allows us to discern a weak absorption at $\sim 27\cm$ indicated by arrow 4 in both polarizations. The strength of resonance 4 is considerably smaller than resonances 1 and 2, which leads to the assignment of resonance 4 as the higher-order $\psi_{\pm 3}$ magnon of the cycloid. The $\psi_{\pm 3}$ mode is separated from the $\psi_{\pm 1}$ modes by $5.3\cm$ in both polarizations $h_1$ and $h_2$, which is compatible with the $\sim7.5-8\cm$ estimate from Raman experiments\cite{singh:252203,cazayous:037601}. The $\psi_{\pm 3}$ mode softens with elevated temperature following the same dependence as the more prominent $\psi_{\pm 1}$ modes and becomes unobservable between 150 and 200 K, most likely due to the reduction in cycloid anharmonicity observed in nuclear magnetic resonance studies\cite{zalessky:547}. 

We now consider resonance 3 that happens at $18.4\cm$ (Fig~\ref{fig:espectra}(a)) and disappears at approximately 150 K, close to the temperature at which a magnetic phase transition occurs, possibly associated with a spin reorientation. Only $\psi_{\pm 1}$ modes are allowed in the $h_1$ polarization, and we cannot assign resonance 3 to the $\psi_0$ mode of the out-of-plane series. The spin reorientation below 140 K may render some of the in-plane modes $\phi_n$ observable in $h_1$ polarization, but that would lead to an expectation of an even stronger resonance at the same frequency in the $h_2$ polarization. Since no such resonance appears in $h_2$ spectra, we conclude that no magnon mode (either $\phi$ or $\psi$) in the de Sousa - Moore description\cite{desousa:012406} is a good candidate for the resonance 3 assignment. One conceivable explanation is a severe modification of the cycloid structure at the 140-K transition, which would allow new lines (in addition to in-plane and out-of-plane cycloid modes) to appear in the magnon spectrum. This possibility is inconsistent with the smooth evolution of the $\psi_{\pm 1}$ modes across the 140-K transition. In addition, no new lines appear at the phase transition in the magnon spectra recorded using Raman scattering\cite{cazayous:037601,singh:252203}. de Sousa and Moore do predict a resonance at a frequency below that of the $\psi_{\pm 1}$ modes: this resonance is excited by the $y$ component of the $ac$ electric field of THz wave and is referred to as electromagnon. The electromagnon selection rule is satisfied in $h_1$ polarization for mode 3, and the frequency of the mode is indeed lower than the frequency of the $\psi_{\pm 1}$ modes. This suggests an assignment of mode 3 as the electromagnon predicted by de Sousa and Moore\cite{desousa:012406}. This assignment leaves open the question of mode 3 disappearance at 150 K. We tentatively attribute changes in electromagnon excitation conditions to the effects of the magnetic phase transition at 140 K.

The observation of magnetic resonance in optical transmission was first reported by Komandin et al.\cite{komandin:734} who studied BFO ceramics. They found four distinct magnetic resonance lines in the 20-$30\cm$ range at low temperature and an additional absorption in the 30-$60\cm$ range with a large dielectric contribution that was identified with an optical phonon. The temperature dependence of the phonon and magnetic resonance modes revealed a coupling between magnetic and lattice excitations. Three magnetic resonance lines found by Komandin et al. coincide in frequency and exhibit the same temperature dependence as the resonances 1, 2, and 4 in both polarizations reported in this paper. The temperature dependence of resonances 1 and 2 in $h_2$ polarization - a doublet at low temperature and a singlet at high temperature (Fig.~\ref{fig:tempfreq}) - closely follows the behavior of the two strongest magnetic resonances reported by Komandin et al., which suggests that those resonances correspond to the $\psi_{+1}$/$\psi_{-1}$ doublet. The third common resonance (our resonance 4) corresponds to the the $\psi_{\pm 3}$ modes. An additional resonance at $24.5\cm$ appeared at low temperature in Komandin's observations and is absent from our spectra. The electromagnon mode at $18.4\cm$ reported here is notably absent from Komandin's results. These differences are likely explained by a combination of two factors, one of which is a sample-to-sample variation. The other is the appearance of additional lines in the spectra (in the work of Komandin et al.\cite{komandin:734} and in Raman spectroscopy\cite{singh:252203, cazayous:037601,rovillain:180411}), which can be accounted for by including single-ion anisotropy and DM terms in the magnetic free energy. As we have shown, a more realistic free energy than the one used by de Sousa and Moore\cite{desousa:012406} is needed for a detailed description of the magnetic resonance spectrum in BFO.

To summarize, we measured THz transmission spectra of BFO at various temperatures and found several absorption resonances which we ascribed to the modes of antiferromagnetic cycloid motion. Magnetic dipole selection rules result in the assignment of the most prominent resonances in $h_1$ and $h_2$ polarizations to the $\psi_{\pm 1}$ modes of the cycloid. The uniform mode $\psi_{0}$ is silent, while the higher-order $\psi_{\pm 3}$ modes exhibit a much smaller oscillator strength. The uniform mode $\phi_{0}$ is allowed in the $h_2$ polarization but is not observed due to its low frequency. A series of magnetic resonance modes at similar frequencies as in our measurements was found in Raman scattering\cite{singh:252203, scott:322203, cazayous:037601, rovillain:180411} and millimeter wave spectroscopy\cite{komandin:734} studies of BFO. The most prominent Raman resonances were found in the $18-30\cm$ range. Some of the Raman results\cite{cazayous:037601} showed several weak absorption lines below $18\cm$ which where attributed to uniform and higher-order modes with n=1,2. The strongest absorption lines between $18$ and $30\cm$ were ascribed to a combined magnon and electromagnon response of the $\psi_3$, $\phi_3$, and $\phi_4$ modes. Remarkably, the observed Raman modes were attributed to excitations with both even and odd indices $n$ in the $\phi_n$ and $\psi_n$ series\cite{cazayous:037601, rovillain:180411}. In our description, the anharmonic spiral is parametrized as $L_z=L_0 \sum_n b_n\cos(nqx)$ with odd integers $n=1,3,...$ The resulting magnetic dipole selection rules for the $\psi_n$ modes allow only odd indices $n$, while for the $\phi_n$ modes only even indices are allowed. We assigned most of the observed absorption lines to magnetic dipole modes $\psi_n$ with $n=1,3$ with the exception of mode 3, for which the de Sousa-Moore electromagnon selection rule is satisfied. Thus, we suggest that mode 3 is an electromagnon due to magnetoelectric coupling between the $E_g$ phonon\cite{lobo:172105} and the $\psi_{\pm 1}$ magnons. Finally, we found a splitting between the $\psi_{+1}$ and $\psi_{-1}$ modes in the $h_1$ polarization and we proposed that it could be due to the modulated DM term in the free energy (Eq. (\ref{eq:dmmod})).

The work at Los Alamos National Laboratory was supported by the LDRD program and by the Center for Integrated Nanotechnologies. The work at Rutgers University was supported by the DOE grant of DE-FG02-07ER46382.


\begin{thebibliography}{31}
\expandafter\ifx\csname natexlab\endcsname\relax\def\natexlab#1{#1}\fi
\expandafter\ifx\csname bibnamefont\endcsname\relax
  \def\bibnamefont#1{#1}\fi
\expandafter\ifx\csname bibfnamefont\endcsname\relax
  \def\bibfnamefont#1{#1}\fi
\expandafter\ifx\csname citenamefont\endcsname\relax
  \def\citenamefont#1{#1}\fi
\expandafter\ifx\csname url\endcsname\relax
  \def\url#1{\texttt{#1}}\fi
\expandafter\ifx\csname urlprefix\endcsname\relax\def\urlprefix{URL }\fi
\providecommand{\bibinfo}[2]{#2}
\providecommand{\eprint}[2][]{\url{#2}}

\bibitem[{\citenamefont{Smith et~al.}(1968)\citenamefont{Smith, Achenbach,
  Gerson, and James}}]{smith:70}
\bibinfo{author}{\bibfnamefont{R.}~\bibnamefont{Smith}},
  \bibinfo{author}{\bibfnamefont{G.}~\bibnamefont{Achenbach}},
  \bibinfo{author}{\bibfnamefont{R.}~\bibnamefont{Gerson}}, \bibnamefont{and}
  \bibinfo{author}{\bibfnamefont{W.}~\bibnamefont{James}}, \bibinfo{journal}{J.
  Appl. Phys.} \textbf{\bibinfo{volume}{39}}, \bibinfo{pages}{70}
  (\bibinfo{year}{1968}).

\bibitem[{\citenamefont{Sosnowska et~al.}(1982)\citenamefont{Sosnowska,
  Peterlin-Neumaier, and Steichele}}]{sosnowska:4835}
\bibinfo{author}{\bibfnamefont{I.}~\bibnamefont{Sosnowska}},
  \bibinfo{author}{\bibfnamefont{T.}~\bibnamefont{Peterlin-Neumaier}},
  \bibnamefont{and}
  \bibinfo{author}{\bibfnamefont{E.}~\bibnamefont{Steichele}},
  \bibinfo{journal}{J. Phys. C} \textbf{\bibinfo{volume}{15}},
  \bibinfo{pages}{4835} (\bibinfo{year}{1982}).

\bibitem[{\citenamefont{Wood and Austin}(1974)}]{wood:303}
\bibinfo{author}{\bibfnamefont{V.}~\bibnamefont{Wood}} \bibnamefont{and}
  \bibinfo{author}{\bibfnamefont{A.}~\bibnamefont{Austin}},
  \bibinfo{journal}{Int. J. Magnetism} \textbf{\bibinfo{volume}{5}},
  \bibinfo{pages}{303} (\bibinfo{year}{1974}).

\bibitem[{\citenamefont{Fiebig}(2005)}]{fiebig:123}
\bibinfo{author}{\bibfnamefont{M.}~\bibnamefont{Fiebig}}, \bibinfo{journal}{J.
  Phys. D: Appl. Phys.} \textbf{\bibinfo{volume}{38}}, \bibinfo{pages}{R123}
  (\bibinfo{year}{2005}).

\bibitem[{\citenamefont{Cheong and Mostovoy}(2007)}]{cheong:13}
\bibinfo{author}{\bibfnamefont{S.-W.} \bibnamefont{Cheong}} \bibnamefont{and}
  \bibinfo{author}{\bibfnamefont{M.}~\bibnamefont{Mostovoy}},
  \bibinfo{journal}{Nature Mat.} \textbf{\bibinfo{volume}{6}},
  \bibinfo{pages}{13} (\bibinfo{year}{2007}).

\bibitem[{\citenamefont{Ramesh and Spaldin}(2007)}]{ramesh:21}
\bibinfo{author}{\bibfnamefont{R.}~\bibnamefont{Ramesh}} \bibnamefont{and}
  \bibinfo{author}{\bibfnamefont{N.}~\bibnamefont{Spaldin}},
  \bibinfo{journal}{Nature Mat.} \textbf{\bibinfo{volume}{6}},
  \bibinfo{pages}{21} (\bibinfo{year}{2007}).

\bibitem[{\citenamefont{Wang et~al.}(2003)}]{wang:1719}
\bibinfo{author}{\bibfnamefont{J.}~\bibnamefont{Wang}} \bibnamefont{et~al.},
  \bibinfo{journal}{Science} \textbf{\bibinfo{volume}{299}},
  \bibinfo{pages}{1719} (\bibinfo{year}{2003}).

\bibitem[{\citenamefont{Shvartsman et~al.}(2007)\citenamefont{Shvartsman,
  Kleeman, Haumont, and Kreisel}}]{shvartsman:172115}
\bibinfo{author}{\bibfnamefont{V.}~\bibnamefont{Shvartsman}},
  \bibinfo{author}{\bibfnamefont{W.}~\bibnamefont{Kleeman}},
  \bibinfo{author}{\bibfnamefont{R.}~\bibnamefont{Haumont}}, \bibnamefont{and}
  \bibinfo{author}{\bibfnamefont{J.}~\bibnamefont{Kreisel}},
  \bibinfo{journal}{Appl. Phys. Lett.} \textbf{\bibinfo{volume}{90}},
  \bibinfo{pages}{1172115} (\bibinfo{year}{2007}).

\bibitem[{\citenamefont{Lebeugle et~al.}(2007)\citenamefont{Lebeugle, Colson,
  Forget, and Viret}}]{lebeugle:22907}
\bibinfo{author}{\bibfnamefont{D.}~\bibnamefont{Lebeugle}},
  \bibinfo{author}{\bibfnamefont{D.}~\bibnamefont{Colson}},
  \bibinfo{author}{\bibfnamefont{A.}~\bibnamefont{Forget}}, \bibnamefont{and}
  \bibinfo{author}{\bibfnamefont{M.}~\bibnamefont{Viret}},
  \bibinfo{journal}{Appl. Phys. Lett.} \textbf{\bibinfo{volume}{91}},
  \bibinfo{pages}{022907} (\bibinfo{year}{2007}).

\bibitem[{\citenamefont{Lebeugle et~al.}(2008)\citenamefont{Lebeugle, Colson,
  Forget, Viret, Bataille, and Goukasov}}]{lebeugle:227602}
\bibinfo{author}{\bibfnamefont{D.}~\bibnamefont{Lebeugle}},
  \bibinfo{author}{\bibfnamefont{D.}~\bibnamefont{Colson}},
  \bibinfo{author}{\bibfnamefont{A.}~\bibnamefont{Forget}},
  \bibinfo{author}{\bibfnamefont{M.}~\bibnamefont{Viret}},
  \bibinfo{author}{\bibfnamefont{A.~M.} \bibnamefont{Bataille}},
  \bibnamefont{and} \bibinfo{author}{\bibfnamefont{A.}~\bibnamefont{Goukasov}},
  \bibinfo{journal}{Phys. Rev. Lett.} \textbf{\bibinfo{volume}{100}},
  \bibinfo{pages}{227602} (\bibinfo{year}{2008}).

\bibitem[{\citenamefont{Lee et~al.}(2008{\natexlab{a}})\citenamefont{Lee,
  Ratcliff, Cheong, and Kiryukhin}}]{lee:192906}
\bibinfo{author}{\bibfnamefont{S.}~\bibnamefont{Lee}},
  \bibinfo{author}{\bibfnamefont{W.}~\bibnamefont{Ratcliff}},
  \bibinfo{author}{\bibfnamefont{S.-W.} \bibnamefont{Cheong}},
  \bibnamefont{and}
  \bibinfo{author}{\bibfnamefont{V.}~\bibnamefont{Kiryukhin}},
  \bibinfo{journal}{Appl. Phys. Lett.} \textbf{\bibinfo{volume}{92}},
  \bibinfo{pages}{192906} (\bibinfo{year}{2008}{\natexlab{a}}).

\bibitem[{\citenamefont{Zhao et~al.}(2006)}]{zhao:823}
\bibinfo{author}{\bibfnamefont{T.}~\bibnamefont{Zhao}} \bibnamefont{et~al.},
  \bibinfo{journal}{Nature Mat.} \textbf{\bibinfo{volume}{5}},
  \bibinfo{pages}{823} (\bibinfo{year}{2006}).

\bibitem[{\citenamefont{Chu et~al.}(2008)}]{chu:478}
\bibinfo{author}{\bibfnamefont{Y.-H.} \bibnamefont{Chu}} \bibnamefont{et~al.},
  \bibinfo{journal}{Nature Mat.} \textbf{\bibinfo{volume}{7}},
  \bibinfo{pages}{478} (\bibinfo{year}{2008}).

\bibitem[{\citenamefont{Kadomtseva et~al.}(2004)\citenamefont{Kadomtseva,
  Zvezdin, Popov, Pyatakov, and Vorob'ev}}]{kadomtseva:571}
\bibinfo{author}{\bibfnamefont{A.}~\bibnamefont{Kadomtseva}},
  \bibinfo{author}{\bibfnamefont{A.}~\bibnamefont{Zvezdin}},
  \bibinfo{author}{\bibfnamefont{Y.}~\bibnamefont{Popov}},
  \bibinfo{author}{\bibfnamefont{A.}~\bibnamefont{Pyatakov}}, \bibnamefont{and}
  \bibinfo{author}{\bibfnamefont{G.}~\bibnamefont{Vorob'ev}},
  \bibinfo{journal}{JETP Lett.} \textbf{\bibinfo{volume}{79}},
  \bibinfo{pages}{571} (\bibinfo{year}{2004}).

\bibitem[{\citenamefont{Kamba et~al.}(2007)\citenamefont{Kamba, Nuzhnyy,
  Savinov, Sebek, Petzelt, Prokleska, Haumont, and Kreisel}}]{kamba:024403}
\bibinfo{author}{\bibfnamefont{S.}~\bibnamefont{Kamba}},
  \bibinfo{author}{\bibfnamefont{D.}~\bibnamefont{Nuzhnyy}},
  \bibinfo{author}{\bibfnamefont{M.}~\bibnamefont{Savinov}},
  \bibinfo{author}{\bibfnamefont{J.}~\bibnamefont{Sebek}},
  \bibinfo{author}{\bibfnamefont{J.}~\bibnamefont{Petzelt}},
  \bibinfo{author}{\bibfnamefont{J.}~\bibnamefont{Prokleska}},
  \bibinfo{author}{\bibfnamefont{R.}~\bibnamefont{Haumont}}, \bibnamefont{and}
  \bibinfo{author}{\bibfnamefont{J.}~\bibnamefont{Kreisel}},
  \bibinfo{journal}{Phys. Rev. B} \textbf{\bibinfo{volume}{75}},
  \bibinfo{pages}{024403} (\bibinfo{year}{2007}).

\bibitem[{\citenamefont{Lobo et~al.}(2007)\citenamefont{Lobo, Moreira,
  Lebeugle, and Colson}}]{lobo:172105}
\bibinfo{author}{\bibfnamefont{R.~P. S.~M.} \bibnamefont{Lobo}},
  \bibinfo{author}{\bibfnamefont{R.~L.} \bibnamefont{Moreira}},
  \bibinfo{author}{\bibfnamefont{D.}~\bibnamefont{Lebeugle}}, \bibnamefont{and}
  \bibinfo{author}{\bibfnamefont{D.}~\bibnamefont{Colson}},
  \bibinfo{journal}{Phys. Rev. B} \textbf{\bibinfo{volume}{76}},
  \bibinfo{pages}{172105} (\bibinfo{year}{2007}).

\bibitem[{\citenamefont{Scott}(1974)}]{scott:83}
\bibinfo{author}{\bibfnamefont{J.}~\bibnamefont{Scott}}, \bibinfo{journal}{Rev.
  Mod. Phys.} \textbf{\bibinfo{volume}{46}}, \bibinfo{pages}{83}
  (\bibinfo{year}{1974}).

\bibitem[{\citenamefont{Foner}(1984)}]{foner}
\bibinfo{author}{\bibfnamefont{S.}~\bibnamefont{Foner}}, in
  \emph{\bibinfo{booktitle}{Magnetism}}, edited by
  \bibinfo{editor}{\bibfnamefont{G.}~\bibnamefont{Rado}} \bibnamefont{and}
  \bibinfo{editor}{\bibfnamefont{H.}~\bibnamefont{Suhl}}
  (\bibinfo{publisher}{Academic Press, New York}, \bibinfo{year}{1984}),
  vol.~\bibinfo{volume}{I}.

\bibitem[{\citenamefont{Aguilar et~al.}(2009)\citenamefont{Aguilar, Mostovoy,
  Sushkov, Zhang, Choi, Cheong, and Drew}}]{valdez:047203}
\bibinfo{author}{\bibfnamefont{R.} \bibnamefont{ValdesAguilar}},
  \bibinfo{author}{\bibfnamefont{M.}~\bibnamefont{Mostovoy}},
  \bibinfo{author}{\bibfnamefont{A.B.}~\bibnamefont{Sushkov}},
  \bibinfo{author}{\bibfnamefont{C.L.}~\bibnamefont{Zhang}},
  \bibinfo{author}{\bibfnamefont{Y.J.}~\bibnamefont{Choi}},
  \bibinfo{author}{\bibfnamefont{S.-W.} \bibnamefont{Cheong}},
  \bibnamefont{and} \bibinfo{author}{\bibfnamefont{H.D.}~\bibnamefont{Drew}},
  \bibinfo{journal}{Phys. Rev. Lett.} \textbf{\bibinfo{volume}{102}},
  \bibinfo{pages}{047203} (\bibinfo{year}{2009}).

\bibitem[{\citenamefont{Pimenov et~al.}(2008)\citenamefont{Pimenov, Loidl,
  Mukhin, Ivanov, and Balbashov}}]{pimenov:014438}
\bibinfo{author}{\bibfnamefont{A.}~\bibnamefont{Pimenov}},
  \bibinfo{author}{\bibfnamefont{A.}~\bibnamefont{Loidl}},
  \bibinfo{author}{\bibfnamefont{A.A.}~\bibnamefont{Mukhin}},
  \bibinfo{author}{\bibfnamefont{V.D.}~\bibnamefont{Travkin}},
  \bibinfo{author}{\bibfnamefont{V.Y.}~\bibnamefont{Ivanov}}, \bibnamefont{and}
  \bibinfo{author}{\bibfnamefont{A.M.}~\bibnamefont{Balbashov}},
  \bibinfo{journal}{Phys. Rev. B} \textbf{\bibinfo{volume}{77}},
  \bibinfo{pages}{014438} (\bibinfo{year}{2008}).

\bibitem[{\citenamefont{de~Sousa and Moore}(2008)}]{desousa:012406}
\bibinfo{author}{\bibfnamefont{R.}~\bibnamefont{de~Sousa}} \bibnamefont{and}
  \bibinfo{author}{\bibfnamefont{J.E.}~\bibnamefont{Moore}},
  \bibinfo{journal}{Phys. Rev. B} \textbf{\bibinfo{volume}{77}},
  \bibinfo{pages}{012406} (\bibinfo{year}{2008}).

\bibitem[{\citenamefont{Lee et~al.}(2008{\natexlab{b}})\citenamefont{Lee, Choi,
  Ratcliff, Erwin, Cheong, and Kiryukhin}}]{lee:100101}
\bibinfo{author}{\bibfnamefont{S.}~\bibnamefont{Lee}},
  \bibinfo{author}{\bibfnamefont{T.}~\bibnamefont{Choi}},
  \bibinfo{author}{\bibfnamefont{W.}~\bibnamefont{Ratcliff}},
  \bibinfo{author}{\bibfnamefont{R.}~\bibnamefont{Erwin}},
  \bibinfo{author}{\bibfnamefont{S.-W.} \bibnamefont{Cheong}},
  \bibnamefont{and}
  \bibinfo{author}{\bibfnamefont{V.}~\bibnamefont{Kiryukhin}},
  \bibinfo{journal}{Phys. Rev. B} \textbf{\bibinfo{volume}{78}},
  \bibinfo{pages}{100101(R)} (\bibinfo{year}{2008}{\natexlab{b}}).

\bibitem[{\citenamefont{Singh et~al.}(2008)\citenamefont{Singh, Katiyar, and
  Scott}}]{singh:252203}
\bibinfo{author}{\bibfnamefont{M.}~\bibnamefont{Singh}},
  \bibinfo{author}{\bibfnamefont{R.}~\bibnamefont{Katiyar}}, \bibnamefont{and}
  \bibinfo{author}{\bibfnamefont{J.}~\bibnamefont{Scott}}, \bibinfo{journal}{J.
  Phys.: Condens. Matter} \textbf{\bibinfo{volume}{20}},
  \bibinfo{pages}{252203} (\bibinfo{year}{2008}).

\bibitem[{\citenamefont{Cazayous et~al.}(2008)\citenamefont{Cazayous, Gallais,
  Sacuto, de~Sousa, Lebeugle, and Colson}}]{cazayous:037601}
\bibinfo{author}{\bibfnamefont{M.}~\bibnamefont{Cazayous}},
  \bibinfo{author}{\bibfnamefont{Y.}~\bibnamefont{Gallais}},
  \bibinfo{author}{\bibfnamefont{A.}~\bibnamefont{Sacuto}},
  \bibinfo{author}{\bibfnamefont{R.}~\bibnamefont{de~Sousa}},
  \bibinfo{author}{\bibfnamefont{D.}~\bibnamefont{Lebeugle}}, \bibnamefont{and}
  \bibinfo{author}{\bibfnamefont{D.}~\bibnamefont{Colson}},
  \bibinfo{journal}{Phys. Rev. Lett.} \textbf{\bibinfo{volume}{101}},
  \bibinfo{pages}{037601} (\bibinfo{year}{2008}).

\bibitem[{\citenamefont{Rovillain et~al.}(2009)\citenamefont{Rovillain,
  Cazayous, Gallais, Sacuto, Lobo, Lebeugle, and Colson}}]{rovillain:180411}
\bibinfo{author}{\bibfnamefont{P.}~\bibnamefont{Rovillain}},
  \bibinfo{author}{\bibfnamefont{M.}~\bibnamefont{Cazayous}},
  \bibinfo{author}{\bibfnamefont{Y.}~\bibnamefont{Gallais}},
  \bibinfo{author}{\bibfnamefont{A.}~\bibnamefont{Sacuto}},
  \bibinfo{author}{\bibfnamefont{R.P.S.M.}~\bibnamefont{Lobo}},
  \bibinfo{author}{\bibfnamefont{D.}~\bibnamefont{Lebeugle}}, \bibnamefont{and}
  \bibinfo{author}{\bibfnamefont{D.}~\bibnamefont{Colson}},
  \bibinfo{journal}{Phys. Rev. B} \textbf{\bibinfo{volume}{79}},
  \bibinfo{pages}{180411(R)} (\bibinfo{year}{2009}).

\bibitem[{\citenamefont{Zalessky et~al.}(2000)\citenamefont{Zalessky, Frolov,
  Khimich, Bush, Pokatilov, and Zvezdin}}]{zalessky:547}
\bibinfo{author}{\bibfnamefont{A.}~\bibnamefont{Zalessky}},
  \bibinfo{author}{\bibfnamefont{A.}~\bibnamefont{Frolov}},
  \bibinfo{author}{\bibfnamefont{T.}~\bibnamefont{Khimich}},
  \bibinfo{author}{\bibfnamefont{A.}~\bibnamefont{Bush}},
  \bibinfo{author}{\bibfnamefont{V.}~\bibnamefont{Pokatilov}},
  \bibnamefont{and} \bibinfo{author}{\bibfnamefont{A.}~\bibnamefont{Zvezdin}},
  \bibinfo{journal}{Europhys. Lett.} \textbf{\bibinfo{volume}{50}},
  \bibinfo{pages}{547} (\bibinfo{year}{2000}).

\bibitem[{\citenamefont{Talbayev et~al.}(2008)\citenamefont{Talbayev, Lee,
  Cheong, and Taylor}}]{talbayev:212906}
\bibinfo{author}{\bibfnamefont{D.}~\bibnamefont{Talbayev}},
  \bibinfo{author}{\bibfnamefont{S.}~\bibnamefont{Lee}},
  \bibinfo{author}{\bibfnamefont{S.-W.} \bibnamefont{Cheong}},
  \bibnamefont{and} \bibinfo{author}{\bibfnamefont{A.}~\bibnamefont{Taylor}},
  \bibinfo{journal}{Appl. Phys. Lett.} \textbf{\bibinfo{volume}{93}},
  \bibinfo{pages}{212906} (\bibinfo{year}{2008}).

\bibitem[{\citenamefont{Komandin et~al.}(2010)\citenamefont{Komandin,
  Torgashev, Volkov, Porodinkov, Spektor, and Bush}}]{komandin:734}
\bibinfo{author}{\bibfnamefont{G.}~\bibnamefont{Komandin}},
  \bibinfo{author}{\bibfnamefont{V.}~\bibnamefont{Torgashev}},
  \bibinfo{author}{\bibfnamefont{A.}~\bibnamefont{Volkov}},
  \bibinfo{author}{\bibfnamefont{O.}~\bibnamefont{Porodinkov}},
  \bibinfo{author}{\bibfnamefont{I.}~\bibnamefont{Spektor}}, \bibnamefont{and}
  \bibinfo{author}{\bibfnamefont{A.}~\bibnamefont{Bush}},
  \bibinfo{journal}{Phys. Solid State} \textbf{\bibinfo{volume}{52}},
  \bibinfo{pages}{734} (\bibinfo{year}{2010}).

\bibitem[{\citenamefont{Scott et~al.}(2008)\citenamefont{Scott, Singh, and
  Katiyar}}]{scott:322203}
\bibinfo{author}{\bibfnamefont{J.}~\bibnamefont{Scott}},
  \bibinfo{author}{\bibfnamefont{M.}~\bibnamefont{Singh}}, \bibnamefont{and}
  \bibinfo{author}{\bibfnamefont{R.}~\bibnamefont{Katiyar}},
  \bibinfo{journal}{J. Phys.: Condens. Matter} \textbf{\bibinfo{volume}{20}},
  \bibinfo{pages}{322203} (\bibinfo{year}{2008}).

\bibitem[{\citenamefont{Xu et~al.}(2009)\citenamefont{Xu, Brinzari, Lee, Chu,
  Martin, Kumar, McGill, Rai, Ramesh, Gopalan et~al.}}]{xu:134425}
\bibinfo{author}{\bibfnamefont{X.}~\bibnamefont{Xu}},
  \bibinfo{author}{\bibfnamefont{T.}~\bibnamefont{Brinzari}},
  \bibinfo{author}{\bibfnamefont{S.}~\bibnamefont{Lee}},
  \bibinfo{author}{\bibfnamefont{Y.}~\bibnamefont{Chu}},
  \bibinfo{author}{\bibfnamefont{L.}~\bibnamefont{Martin}},
  \bibinfo{author}{\bibfnamefont{A.}~\bibnamefont{Kumar}},
  \bibinfo{author}{\bibfnamefont{S.}~\bibnamefont{McGill}},
  \bibinfo{author}{\bibfnamefont{R.}~\bibnamefont{Rai}},
  \bibinfo{author}{\bibfnamefont{R.}~\bibnamefont{Ramesh}},
  \bibinfo{author}{\bibfnamefont{V.}~\bibnamefont{Gopalan}},
  \bibnamefont{et~al.}, \bibinfo{journal}{Phys. Rev. B}
  \textbf{\bibinfo{volume}{79}}, \bibinfo{pages}{134425}
  (\bibinfo{year}{2009}).

\bibitem[{\citenamefont{Ruette et~al.}(2004)\citenamefont{Ruette, Zvyagin,
  Pyatakov, Bush, Li, Belotelov, Zvezdin, and Viehland}}]{ruette:064114}
\bibinfo{author}{\bibfnamefont{B.}~\bibnamefont{Ruette}},
  \bibinfo{author}{\bibfnamefont{S.}~\bibnamefont{Zvyagin}},
  \bibinfo{author}{\bibfnamefont{A.P.}~\bibnamefont{Pyatakov}},
  \bibinfo{author}{\bibfnamefont{A.}~\bibnamefont{Bush}},
  \bibinfo{author}{\bibfnamefont{J.F.}~\bibnamefont{Li}},
  \bibinfo{author}{\bibfnamefont{V.I.}~\bibnamefont{Belotelov}},
  \bibinfo{author}{\bibfnamefont{A.K.}~\bibnamefont{Zvezdin}}, \bibnamefont{and}
  \bibinfo{author}{\bibfnamefont{D.}~\bibnamefont{Viehland}},
  \bibinfo{journal}{Phys. Rev. B} \textbf{\bibinfo{volume}{69}},
  \bibinfo{pages}{064114} (\bibinfo{year}{2004}).

\end{thebibliography}

\newpage

\end{document}